\documentclass[12pt]{iopart}

\usepackage{iopams}       
\usepackage{graphicx}     
\usepackage{longtable}    
\usepackage{cite}         

\newcommand*{\half}{\frac{1}{2}}

\begin{document}

\title{Studying springs in series using a single spring}
\author{Juan D. Serna$^1$ and Amitabh Joshi$^2$}
\address{$^1$ School of Mathematical and Natural Sciences,
              University of Arkansas at Monticello, Monticello, AR 71656, USA}
\address{$^2$ Department of Physics, Eastern Illinois University,
              Charleston, IL 61920, USA}
\ead{\mailto{serna@uamont.edu} and \mailto{ajoshi@eiu.edu}}

\begin{abstract}
Springs are used for a wide range of applications in physics and engineering.
Possibly, one of its most common uses is to study the nature of restoring forces
in oscillatory systems. While experiments that verify the Hooke's law using
springs are abundant in the physics literature, those that explore the
combination of several springs together are very rare. In this paper, an
experiment designed to study the static properties of a combination of springs
in series using only one single spring is presented. Paint marks placed on
the coils of the spring allowed us to divide it into segments, and considered it
as a collection of springs connected in series. The validity of Hooke's law for
the system and the relationship between the spring constant of the segments with
the spring constant of the entire spring is verified experimentally. The easy
setup, accurate results, and educational benefits make this experiment
attractive and useful for high school and first-year college students.
\end{abstract}

\maketitle


\section{Introduction}

Restoring forces play a very fundamental role in the study of vibrations of
mechanical systems. If a system is moved from its equilibrium position, a
restoring force will tend to bring the system back toward equilibrium. For
decades, if not centuries, springs have been used as the most common example of
this type of mechanical system, and have been used extensively to study the
nature of restoring forces. In fact, the use of springs to demonstrate the
Hooke's law is an integral part of every elementary physics lab. However, and
despite the fact that many papers have been written on this topic, and several
experiments designed to verify that the extension of a spring is, in most cases,
directly proportional to the force exerted on
it~\cite{Mills:404,Cushing:925,Easton:494,Hmurcik:135,
Sherfinski:552,Glaser:164,Menz:483,Wagner:566,Souza:35,Struganova:516,
Freeman:224,Euler:57}, not much has been written about experiments concerning
springs connected in series. Perhaps one of the most common reasons why little
attention has been paid to this topic is the fact that a mathematical
description of the physical behaviour of springs in series can be derived
easily~\cite{Gilbert:430}. Most of the textbooks in fundamental physics rarely
discuss the topic of springs in series, and they just leave it as an end of the
chapter problem for the student~\cite{Giancoli,Serway}.

One question that often arises from spring experiments is, ``If a uniform spring
is cut into two or three segments, what is the spring constant of each
segment?'' This paper describes a simple experiment to study the combination of
springs in series using only \textit{one} single spring. The goal is to prove
experimentally that Hooke's law is satisfied not only by each individual spring
of the series, but also by the \textit{combination} of springs as a whole. To
make the experiment effective and easy to perform, first we avoid cutting a
brand new spring into pieces, which is nothing but a waste of resources and
equipment misuse; second, we avoid combining in series several springs with
dissimilar characteristics. This actually would not only introduce additional
difficulties in the physical analysis of the problem (different mass densities
of the springs), but it would also be a source of random error, since the points
at which the springs join do not form coils and the segment elongations might
not be recorded with accuracy. Moreover, contact forces (friction) at these
points might affect the position readings, as well. Instead, we decide just to
use one single spring with paint marks placed on the coils that allow us to
divide it into different segments, and consider it as a collection of springs
connected in series. Then the static Hooke's exercise is carried out on the
spring to observe how each segment elongates under a suspended mass.

In the experiment, two different scenarios are examined: the mass-spring system
with an ideal massless spring, and the realistic case of a spring whose mass is
comparable to the hanging mass. The graphical representation of force against
elongation, used to obtain the spring constant of each individual segment,
shows, in excellent agreement with the theoretical predictions, that the inverse
of the spring constant of the entire spring equals the addition of the
reciprocals of the spring constants of each individual segment. Furthermore, the
experimental results allow us to verify that the ratio of the spring constant of
a segment to the spring constant of the entire spring equals the ratio of the
total number of coils of the spring to the number of coils of the segment.

The experiment discussed in this article has some educational benefits that may
make it attractive for a high school or a first-year college laboratory: It is
easy to perform by students, makes use of only one spring for the investigation,
helps students to develop measuring skills, encourages students to use
computational tools to do linear regression and propagation of error analysis,
helps to understand how springs work using the relationship between the spring
constant and the number of coils, complements the traditional static Hooke's law
experiment with the study of combinations of springs in series, and explores the
contribution of the spring mass to the total elongation of the spring.

\section{The model}

When a spring is stretched, it resists deformation with a force proportional to
the amount of elongation. If the elongation is not too large, this can be
expressed by the approximate relation $F = -k\,x$, where $F$ is the restoring
force, $k$ is the spring constant, and $x$ is the elongation (displacement of
the end of the spring from its equilibrium position)~\cite{Symon}. Because most
of the springs available today are \textit{preloaded}, that is, when in the
relaxed position, almost all of the adjacent coils of the helix are in contact,
application of only a minimum amount of force (weight) is necessary to stretch
the spring to a position where all of the coils are separated from each
other~\cite{Glanz:1091,Prior:601,Froehle:368}. At this new position, the spring
response is linear, and Hooke's law is satisfied.

It is not difficult to show that, when two or more springs are combined in
series (one after another), the resulting combination has a spring constant less
than any of the component springs. In fact, if $p$ ideal springs are connected
in sequence, the expression
\begin{equation}
\frac{1}{k} = \sum_{i=1}^p \frac{1}{k_i}
\label{Eq:1/k}
\end{equation}
relates the spring constant $k$ of the combination with the spring constant
$k_i$ of each individual segment. In general, for a cylindrical spring of
spring constant $k$ having $N$ coils, which is divided into smaller segments,
having $n_i$ coils, the spring constant of each segment can be written as
\begin{equation}
k_i = \frac{N}{n_i} k\,.
\label{Eq:ki}
\end{equation}
Excluding the effects of the material from which a spring is made, the diameter
of the wire and the radius of the coils, this equation expresses the fact that
the spring constant $k$ is a parameter that depends on the number of coils $N$
in a spring, but not on the way in which the coils are wound (i.e. tightly or
loosely)~\cite{Gilbert:430}.

In an early paper, Galloni and Kohen~\cite{Galloni:1076} showed that, under
\textit{static} conditions, the elongation sustained by a non-null mass spring
is equivalent to assuming that the spring is massless and a fraction of one-half
of the spring mass should be added to the hanging mass. That is, if a spring of
mass $m_{\mathrm{s}}$ and relaxed length $l$ (neither stretched nor compressed)
is suspended vertically from one end in the Earth's gravitational field, the
mass per unit length becomes a function of the position, and the spring
stretches \textit{non-uniformly} to a new length $l' = l + \Delta l$. When a
mass $m$ is hung from the end of the spring, the total elongation $\Delta l$ is
found to be
\begin{equation}
\Delta l = \int_0^l \xi(x)\,\rmd x = \frac{(m + \half m_{\mathrm{s}})\,g}{k}\,,
\label{Eq:Dl1}
\end{equation}
where
\begin{equation}
\xi(x) = \frac{m + m_{\mathrm{s}}(l-x)/l}{k\,l}\,g
\label{Eq:xi}
\end{equation}
is the \textit{dimensionless elongation factor} of the element of length between
$x$ and $x + \rmd x$, and $g$ is the acceleration due to gravity. An important
number of papers dealing with the static and dynamic effects of the spring mass
have been written in the physics education literature. Expressions for the
spring elongation as a function of the $n$th coil and the mass per unit length
of the spring have also been derived~\cite{Edwards:445,Heard:1102,Lancaster:217,
Mak:994,French:244,Hosken:327,Ruby:140,Sawicki:276,Ruby:324,Toepker:16,
Newburgh:586,Bowen:1145,Christensen:818,Rodriguez:100,Gluck:178,Essen:603}.

\section{The Experiment}

We want to show that, with just \textit{one} single spring, it is possible to
confirm experimentally the validity of equations~\eref{Eq:1/k}
and~\eref{Eq:ki}. This approach differs from Souza's work~\cite{Souza:35} in
that the constants $k_i$ are determined from the same single spring, and there
is no need of cutting the spring into pieces; and from the standard experiment
in which more than one spring is required.

\begin{figure}
\centering
\includegraphics[scale=1.2]{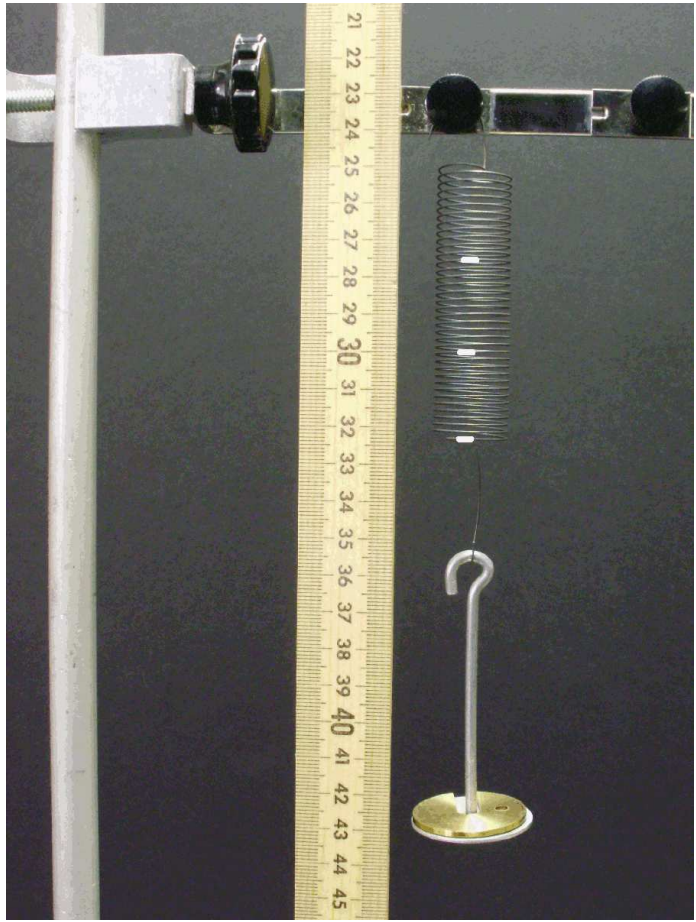}
\caption{\label{Fig1} Experimental setup to determine the spring constants of
springs in series. A soft spring of mass $m_{\mathrm{s}} = 4.43 \pm
0.01\,\mbox{g}$, radius $r = 1.0 \pm 0.1\,\mbox{cm}$, and $n=36$ turns, is
divided into segments by using paint marks.}
\end{figure}

\begin{figure}
\centering
\includegraphics[scale=1.2]{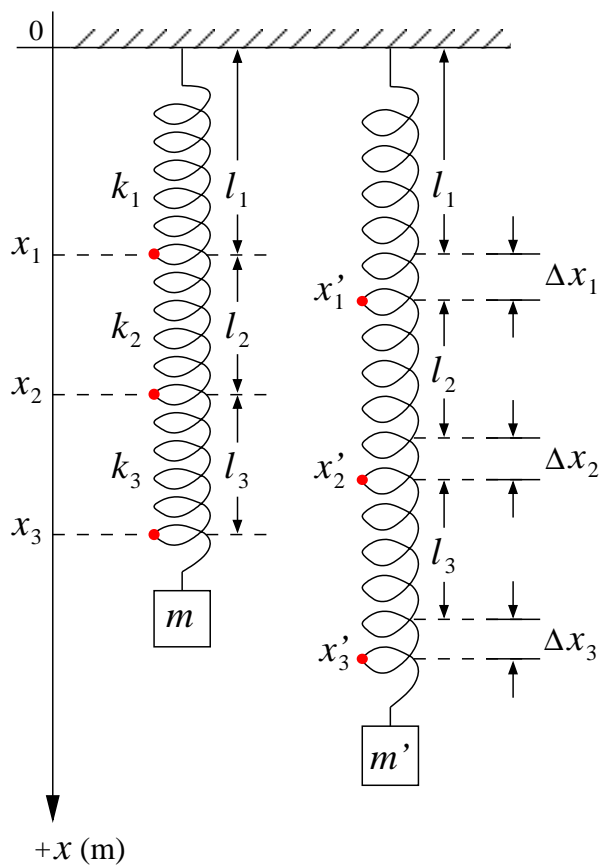}
\caption{\label{Fig2} Schematic of the mass-spring system. An initial mass $m$
is suspended from the spring to bring it into its linear region. $l_i$ is the
initial length of the $i$th spring segment with spring constant $k_i$
($i=1,2,3$). An additional mass $m'$ suspended from the spring elongates each
segment by a distance $\Delta x_i$.}
\end{figure}

A soft spring is \textit{divided} into three separate segments by placing a
paint mark at selected points along its surface (see~\fref{Fig1}). These points
are chosen by counting a certain number of coils for each individual segment
such that the original spring is now composed of three marked springs connected
in series, with each segment represented by an index $i$ (with $i=1,2,3$), and
consisting of $n_i$ coils. An initial mass $m$ is suspended from the spring to
stretch it into its \textit{linear} region, where the equation $F_i=-k_i\Delta
x_i$ is satisfied by each segment. Once the spring is brought into this region,
the traditional static Hooke's law experiment is performed for several different
suspended masses, ranging from $1.0$ to $50.0\,\mbox{g}$. The initial positions
of the marked points $x_i$ are then used to measure the \textit{relative}
displacement (elongation) of each segment after they are stretched by the
additional masses suspended from the spring~(\fref{Fig2}). The displacements are
determined by the equations
\begin{equation}
\Delta x_i = (x'_i - x'_{i-1}) - l_i\,,
\label{Eq:Dxi1}
\end{equation}
where the primed variables $x'_i$ represent the new positions of the marked
points, $l_i = x_i - x_{i-1}$ are the initial lengths of the spring segments,
and $x_0 = 0$, by definition. Representative graphs used to determine the spring
constant of each segment are shown in figures~\ref{Fig3},~\ref{Fig4},
and~\ref{Fig5}.

\section{Dealing with the effective mass}

As pointed out by some
authors~\cite{Galloni:1076,Mak:994,French:244,Sawicki:276,Newburgh:586}, it is
important to note that there is a difference in the total mass hanging from each
segment of the spring. The reason is that each segment supports not only the
mass of the segments below it, but also the mass attached to the end of the
spring. For example, if a spring of mass $m_i$ is divided into three
\textit{identical} segments, and a mass $m$ is suspended from the end of it, the
total mass $M_1$ hanging from the first segment becomes $m +
\frac{2}{3}m_{\mathrm{s}}$. Similarly, for the second and third segments, the
total masses turn out to be $M_2 = m + \frac{1}{3}m_{\mathrm{s}}$ and $M_3 = m
$, respectively. However, in a more realistic scenario, the mass of the spring
and its effect on the elongation of the segments must be considered, and
equation~\eref{Eq:Dl1} should be incorporated into the calculations. Therefore,
for each individual segment, the elongation should be given by
\begin{equation}
\Delta x_i = \frac{(M_i + \half m_i)\,g}{k_i},
\label{Eq:Dxi2}
\end{equation}
where $m_i$ is the mass of the $i$th segment, $M_i$ is its corresponding total
hanging mass, and $k_i$ is the segment's spring constant. Consequently, for the
spring divided into three identical segments ($m_i = \frac{1}{3}
m_{\mathrm{s}}$), the total masses hanging from the first, second and third
segments are now $m + \frac{5}{6} m_{\mathrm{s}}$, $m + \frac{1}{2}
m_{\mathrm{s}}$ and $m + \frac{1}{6} m_{\mathrm{s}}$, respectively. This can be
explained by the following simple consideration: If a mass $m$ is attached to
the end of a spring of length $l$ and spring constant $k$, for three identical
segments with elongations $\Delta l_1$, $\Delta l_2$, and $\Delta l_3$, the
total spring elongation is given by
\begin{eqnarray}
\Delta l &= \Delta l_1 + \Delta l_2 + \Delta l_3 \nonumber
\\[10pt]
&= \int_0^{\frac{l}{3}} \xi(x)\,\rmd x +
   \int_{\frac{l}{3}}^{\frac{2l}{3}} \xi(x)\,\rmd x +
   \int_{\frac{2l}{3}}^l \xi(x)\,\rmd x \nonumber
\\[10pt]
&= \frac{(m + \frac{5}{6} m_{\mathrm{s}})\,g}{3\,k} +
   \frac{(m + \half m_{\mathrm{s}})\,g}{3\,k} +
   \frac{(m + \frac{1}{6} m_{\mathrm{s}})\,g}{3\,k} \nonumber
\\[10pt]
&= \frac{(m + \half m_{\mathrm{s}})\,g}{k}\,.
\label{Eq:Dl2}
\end{eqnarray}
As expected, equation~\eref{Eq:Dl2} is in agreement with
equation~\eref{Eq:Dl1}, and reveals the contribution of the mass of each
individual segment to the total elongation of the spring. It is also observed
from this equation that
\begin{equation}
\Delta l_1 - \Delta l_2 = \Delta l_2 - \Delta l_3 =
\frac{(\frac{1}{3} m_{\mathrm{s}})g}{3\,k} = \mbox{const.}
\label{Eq:Dl123}
\end{equation}
As we know, $\frac{1}{3} m_{\mathrm{s}}$ is the mass of each identical segment,
and $k_1 = k_2 = k_3 = 3\,k$  is the spring constant for each. Therefore, the
spring stretches non-uniformly under its own weight, but uniformly under the
external load, as it was also indicated by Sawicky~\cite{Sawicki:276}.

\section{Results and Discussion}

Two particular cases were studied in this experiment. First, we considered a
spring-mass system in which the spring mass was small compared with the hanging
mass, and so it was ignored. In the second case, the spring mass was comparable
with the hanging mass and included in the calculations.

We started with a configuration of three approximately \textit{identical} spring
segments connected in series; each segment having $12$ coils
($n_1 = n_2 = n_3 = 12$)~\footnote{Although the three segments had the same
number of coils, the first and third segments had an additional portion of wire
where the spring was attached and the masses suspended. This added extra mass to
these segments, making them slightly different from each other and from the
second segment.} When the spring was stretched by different weights, the
elongation of the segments increased linearly, as expected from Hooke's law.
Within the experimental error, each segment experienced the same displacement,
as predicted by~\eref{Eq:Dl123}. An example of experimental data obtained is
shown in~\tref{Table01}.

\begin{figure}
\centering
\includegraphics[scale=1.2]{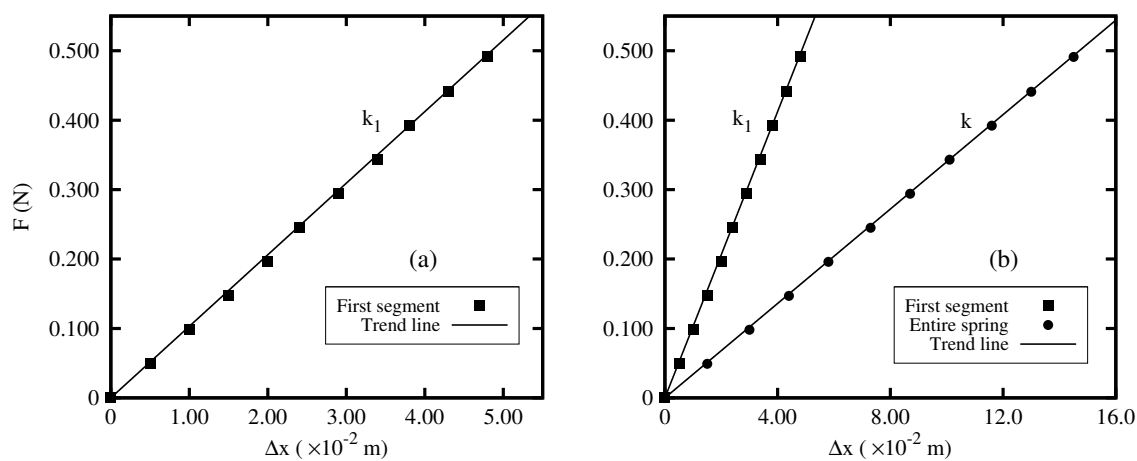}
\caption{\label{Fig3} Applied force as a function of the displacement for the
first spring segment and the total spring. The spring was considered massless
and divided into three \textit{identical} segments ($n_1=n_2=n_3=12$). (a) The
spring constant of the first segment, $k_1=10.3\pm0.1\,\mbox{N/m}$, was obtained
from the slope of the trend line. (b) A comparison between elongations of the
first segment and total spring. Here, $k=3.40\pm0.01\,\mbox{N/m}$. The spring
constants $k_2=10.1\pm0.1\,\mbox{N/m}$ and $k_3=10.2\pm0.1\,\mbox{N/m}$ were
also calculated. It can be observed that $k_i = 3\,k$, as predicted by
equation~\eref{Eq:ki}.}
\end{figure}

Simple linear regression was used to determine the slope of each trend line
fitting the data points of the force versus displacement graphs. \Fref{Fig3}(a)
clearly shows the linear response of the first segment of the spring, with a
resulting spring constant of $k_1=10.3\,\pm\,0.1\,\mbox{N/m}$. A similar
behaviour was observed for the second and third segments, with spring constants
$k_2=10.1\,\pm\,0.1\,\mbox{N/m}$, and $k_3=10.2\,\pm\,0.1\,\mbox{N/m}$,
respectively. For the entire spring, the spring constant was
$k=3.40\pm\,0.01\,\mbox{N/m}$, as shown in~\fref{Fig3}(b). The uncertainties in
the spring constants were calculated using the \textit{correlation coefficient}
$R$ of the linear regressions, as explained in Higbie's paper ``Uncertainty in
the linear regression slope''~\cite{Higbie:184}. Comparing the spring constant
of each segment with that for the total spring, we obtained that $k_1=3.03\,k$,
$k_2=2.97\,k$ and $k_3=3.00\,k$. As predicted by~\eref{Eq:ki}, each segment had
a spring constant three times larger than the resulting combination of the
segments in series, that is, $k_i = 3\,k$.

The reason why the uncertainty in the spring constant of the entire spring is
smaller than the corresponding spring constants of the segments may be explained
by the fact that the displacements of the spring as a whole have smaller
``relative errors'' than those of the individual segments. \Tref{Table01} shows
that, whereas the displacements of the individual segments $\Delta x_i$ are in
the same order of magnitude that the uncertainty in the measurement of the
elongation ($\pm 0.002\,\mbox{m}$), the displacements of the whole spring
$\Delta x_{\mathrm{s}}$ are much bigger compared with this uncertainty.

\begin{figure}
\centering
\includegraphics[scale=1.2]{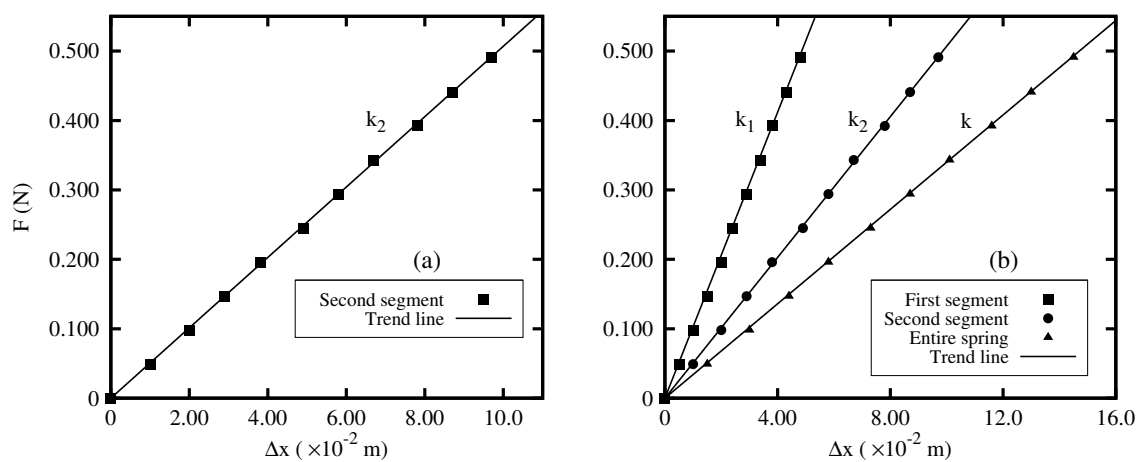}
\caption{\label{Fig4} Applied force as a function of the displacement for the
first and second spring segments, and the total spring. The spring was
considered massless and divided into two \textit{non-identical} segments ($n_2=2
\,n_1=24$). (a) The spring constant of the second segment is
$k_2=5.07\pm0.03\,\mbox{N/m}$. (b) A comparison between elongations of the first
and second segments with the total spring. $k_1=10.3\pm0.1\,\mbox{N/m}$ and
$k=3.40\pm0.01\,\mbox{N/m}$. Here, $k_1 = 3\,k$ and $k_2 = \frac{3}{2}\,k$.}
\end{figure}

\begin{figure}
\centering
\includegraphics[scale=1.2]{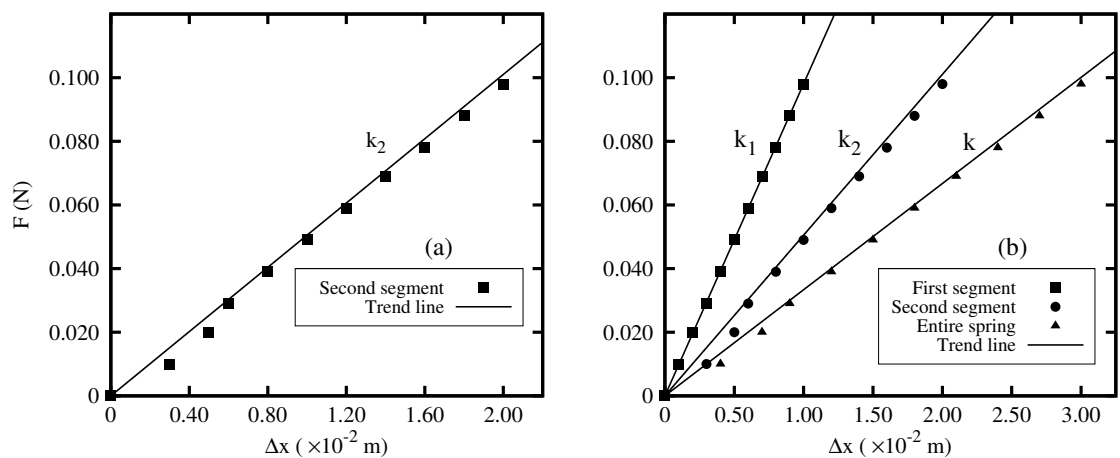}
\caption{\label{Fig5} Applied force as a function of the displacement for the
first and second spring segments, and the total spring. The mass of the spring
was included in the experimental calculations, and the spring divided into two
\textit{non-identical} segments ($n_2=2 n_1=24$). (a) The spring constants of
the segments were calculated with the corrections to the mass. The second
segment has a spring constant of $k_2=5.05\pm0.09\,\mbox{N/m}$. (b) Differences
between the spring elongations of the two segments and the total spring are
shown. Here, $k=3.34\pm0.04\,\mbox{N/m}$.}
\end{figure}

We next considered a configuration of two spring segments connected in series
with $12$ and $24$ coils, respectively ($n_1 = 12$, $n_2 = 24$). \Fref{Fig4}(a)
shows a graph of force against elongation for the second segment of the spring.
We obtained $k_2=5.07\,\pm\,0.03\,\mbox{N/m}$ using linear regression. For the
first segment and the entire spring, the spring constants were
$k_1=10.3\,\pm\,0.1\,\mbox{N/m}$ and $k=3.40\,\pm\,0.01\,\mbox{N/m}$,
respectively, as shown in~\fref{Fig4}(b). Then, we certainly observed that $k_1
= 3.03\,k$ and $k_2 = 1.49\,k$. Once again, these experimental results
proved equation~\eref{Eq:ki} correct ($k_1 = 3\,k$ and $k_2 = \frac{3}{2}\,k$).

We finally considered the same two spring configuration as above, but unlike the
previous trial, this time the spring mass ($4.5 \pm 0.1\,\mbox{g}$) was included
in the experimental calculations. Figures~\ref{Fig5}(a)--(b) show results for
the two spring segments, including spring masses, connected in series ($n_1 =
12$, $n_2 = 24$). Using this method, the spring constant for the whole spring
was found to be slightly different from that obtained when the spring was
assumed ideal (massless). This difference may be explained by the corrections
made to the total mass as given by~\eref{Eq:Dl2}. The spring constants
obtained for the segments were $k_1 = 2.94\,k$ and $k_2 = 1.51\,k$ with $k =
3.34 \pm 0.04\,\mbox{N/m}$ for the entire spring. These experimental results
were also consistent with equation~\eref{Eq:ki}. The experimental data obtained
is shown in~\tref{Table02}.

When the experiment was performed by the students, measuring the positions of
the paint marks on the spring when it was stretched, perhaps represented the
most difficult part of the activity. Every time that an extra weight was added
to the end of the spring, the starting point of each individual segment changed
its position. For the students, keeping track of these new positions was a
laborious task. Most of the experimental systematic error came from this portion
of the activity. To obtain the elongation of the segments, using
equation~\eref{Eq:Dxi1} substantially facilitated the calculation and tabulation
of the data for its posterior analysis. The use of computational tools
(spreadsheets) to do the linear regression, also considerably simplified the
calculations.

\section{Conclusions}

In this work, we studied experimentally the validity of the static Hooke's law
for a system of springs connected in series using a simple single-spring scheme
to represent the combination of springs. We also verified experimentally the
fact that the reciprocal of the spring constant of the entire spring equals the
addition of the reciprocal of the spring constant of each segment by including
well-known corrections (due to the finite mass of the spring) to the total
hanging mass. Our results quantitatively show the validity of Hooke's law for
combinations of springs in series [equation~\eref{Eq:1/k}], as well as the
dependence of the spring constant on the number of coils in a spring
[equation~\eref{Eq:ki}]. The experimental results were in excellent agreement,
within the standard error, with those predicted by theory.

The experiment is designed to provide several educational benefits to the
students, like helping to develop measuring skills, encouraging the use of
computational tools to perform linear regression and error propagation analysis,
and stimulating the creativity and logical thinking by exploring Hooke's law in
a combined system of \textit{springs in series} simulated by a \textit{single}
spring. Because of it easy setup, this experiment is easy to adopt in any high
school or undergraduate physics laboratory, and can be extended to any number of
segments within the same spring such that all segments represent a combination
of springs in series.

\ack
The authors gratefully acknowledge the School of Mathematical and Natural
Sciences at the University of Arkansas-Monticello (\#11-2225-5-M00) and the
Department of Physics at Eastern Illinois University for providing funding and
support for this work. Comments on earlier versions of the paper were gratefully
received from Carol Trana. The authors are also indebted to the anonymous
referee for the valuable comments and suggestions made.


\newpage
\Tables

\Table{\label{Table01} Spring divided into three \textit{identical} segments
($n_1 = n_2 = n_3 = 12$). The $\Delta x_i$ corresponds to the relative
displacement of the $i$th segment (with $i=1,2,3$), and $\Delta x_{\mathrm{s}}$
represents the relative displacement of the entire spring.}
\br
\centre{4}{Displacement} & & \centre{1}{Force} \\
\ms
\centre{4}{($\pm\,0.002\,\mbox{m}$)} & & \centre{1}{$(\pm\,0.001\,\mbox{N}$)} \\
\mr
$\Delta x_1$ & $\Delta x_2$ & $\Delta x_3$ & $\Delta x_{\mathrm{s}}$ & & $F$ \\
\mr
0.000 & 0.000 & 0.000 & 0.000 & & 0.000 \\
0.005 & 0.005 & 0.005 & 0.015 & & 0.049 \\
0.010 & 0.010 & 0.010 & 0.030 & & 0.098 \\
0.015 & 0.014 & 0.015 & 0.044 & & 0.147 \\
0.020 & 0.019 & 0.019 & 0.058 & & 0.196 \\
0.024 & 0.024 & 0.025 & 0.073 & & 0.245 \\
0.029 & 0.029 & 0.029 & 0.087 & & 0.294 \\
0.034 & 0.034 & 0.033 & 0.101 & & 0.343 \\
0.038 & 0.039 & 0.039 & 0.116 & & 0.392 \\
0.043 & 0.044 & 0.043 & 0.130 & & 0.441 \\
0.048 & 0.048 & 0.049 & 0.145 & & 0.491 \\
\br
\endTable

\Table{\label{Table02} Spring divided into \textit{non-identical} segments
($n_1 = 12$, $n_2 = 24$). The mass of the spring was included in the
experimental calculations. The $\Delta x_i$ corresponds to the relative
displacement of the $i$th segment (with $i=1,2$), and $\Delta x_{\mathrm{s}}$
represents the relative displacement of the entire spring.}
\br
\centre{3}{Displacement} & & \centre{1}{Force} \\
\ms
\centre{3}{($\pm\,0.002\,\mbox{m}$)} & & \centre{1}{$(\pm\,0.001\,\mbox{N}$)} \\
\mr
$\Delta x_1$ & $\Delta x_2$ & $\Delta x_{\mathrm{s}}$ & & $F$ \\
\mr
0.000 & 0.000 & 0.000 & & 0.000 \\
0.001 & 0.003 & 0.004 & & 0.010 \\
0.002 & 0.005 & 0.007 & & 0.020 \\
0.003 & 0.006 & 0.009 & & 0.029 \\
0.004 & 0.008 & 0.012 & & 0.039 \\
0.005 & 0.010 & 0.015 & & 0.049 \\
0.006 & 0.012 & 0.018 & & 0.059 \\
0.007 & 0.014 & 0.021 & & 0.069 \\
0.008 & 0.016 & 0.024 & & 0.078 \\
0.009 & 0.018 & 0.027 & & 0.088 \\
0.010 & 0.020 & 0.030 & & 0.098 \\
\br
\endTable

\newpage
\Figures

\newpage

\newpage

\newpage

\end{document}